\documentclass[10pt,twocolumn,letterpaper]{article}

\usepackage{cvpr}              

\usepackage[accsupp]{axessibility}
\usepackage{graphicx}
\usepackage{amsmath}
\usepackage{amssymb}
\usepackage{booktabs}
\usepackage{pifont}
\usepackage{algorithm}
\usepackage{algpseudocode}
\usepackage{enumitem}
\usepackage{float}
\usepackage{tabularx}
\usepackage{adjustbox}
\usepackage{multirow}
\usepackage{multicol}
\usepackage{color, colortbl}
\usepackage{xcolor}
\usepackage{colortbl}
\usepackage{cellspace}

\definecolor{Gray}{gray}{0.9}
\definecolor{LightCyan}{rgb}{0.88,1,1}
\definecolor{sh_gray}{rgb}{0.84,0.84,0.84}
\definecolor{sh_gray2}{rgb}{1,0.89,0.75}
\definecolor{color3}{rgb}{0.95,0.95,0.95}
\definecolor{color4}{rgb}{0.96,0.96,0.86}
\definecolor{color5}{rgb}{0.90,0.90,0.90}

\newcommand{\cmark}{\ding{51}}%
\newcommand{\xmark}{\ding{55}}%
\algnewcommand{\Inputs}[1]{%
  \State \textbf{Inputs:}
  \Statex \hspace*{\algorithmicindent}\parbox[t]{.8\linewidth}{\raggedright #1}
}
\algnewcommand{\Initialize}[1]{%
  \State \textbf{Initialize:}
  \Statex \hspace*{\algorithmicindent}\parbox[t]{.8\linewidth}{\raggedright #1}
}

\usepackage[pagebackref,breaklinks,colorlinks]{hyperref}

\usepackage[capitalize]{cleveref}
\crefname{section}{Sec.}{Secs.}
\Crefname{section}{Section}{Sections}
\Crefname{table}{Table}{Tables}
\crefname{table}{Tab.}{Tabs.}

\def\cvprPaperID{} 
\def\confName{CVPR}
\def\confYear{2022}

\begin{document}

\title{Conformer and Blind Noisy Students for Improved Image Quality Assessment}

\author{Marcos V. Conde, Maxime Burchi, Radu Timofte\\
Computer Vision Lab, Institute of Computer Science, University of Würzburg, Germany\\
{\tt\small \{marcos.conde-osorio,maxime.burchi,radu.timofte\}@uni-wuerzburg.de}
}

\maketitle


\begin{abstract}
Generative models for image restoration, enhancement, and generation have significantly improved the quality of the generated images. Surprisingly, these models produce more pleasant images to the human eye than other methods, yet, they may get a lower perceptual quality score using traditional perceptual quality metrics such as PSNR or SSIM. Therefore, it is necessary to develop a quantitative metric to reflect the performance of new algorithms, which should be well-aligned with the person's mean opinion score (MOS).

Learning-based approaches for perceptual image quality assessment (IQA) usually require both the distorted and reference image for measuring the perceptual quality accurately. However, commonly only the distorted or generated image is available. In this work, we explore the performance of transformer-based full-reference IQA models. We also propose a method for IQA based on semi-supervised knowledge distillation from full-reference teacher models into blind student models using noisy pseudo-labeled data.

Our approaches achieved competitive results on the NTIRE 2022 Perceptual Image Quality Assessment Challenge: our full-reference model was ranked 4th, and our blind noisy student was ranked 3rd among 70 participants, each in their respective track.
\url{https://github.com/burchim/IQA-Conformer-BNS}.
\end{abstract}


\section{Introduction}
\label{sec:intro}

\begin{figure*}[ht]
    \centering
    \setlength{\tabcolsep}{2.0pt}
    \begin{tabular}{ccccc}
    \includegraphics[width=0.19\linewidth]{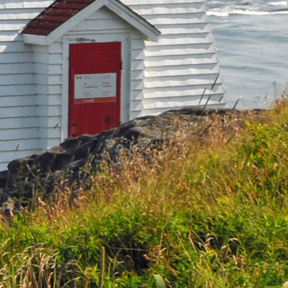} &
    \includegraphics[width=0.19\linewidth]{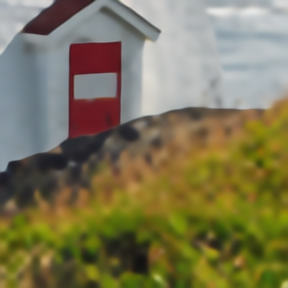} &
    \includegraphics[width=0.19\linewidth]{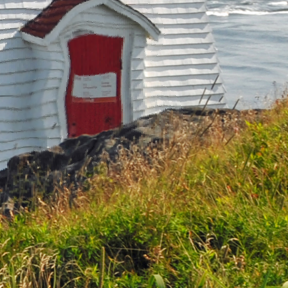} &
    \includegraphics[width=0.19\linewidth]{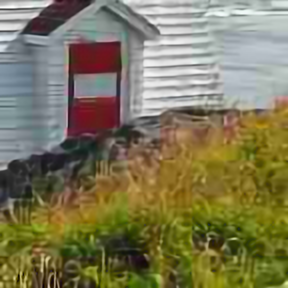} & 
    \includegraphics[width=0.19\linewidth]{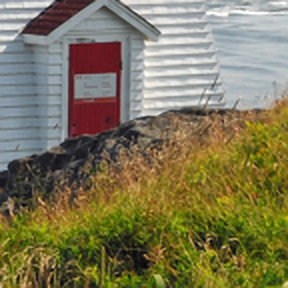}
    \tabularnewline
    Reference & 
    (a) PSNR 29.22 & (b) PSNR 34.65 & (c) PSNR 29.43 & (d) PSNR 31.66 \tabularnewline
    & SSIM 0.37 & SSIM 0.82 & SSIM 0.62 & SSIM 0.90 \tabularnewline
    & MOS 1222.33  & MOS 1318.83 & MOS 1390.17 & MOS 1608.99
    \tabularnewline
    \end{tabular}
    \caption{Training samples from the PIPAL~\cite{pipal, gu2021ntire}. As we can see, ranking the images by their perceptual quality depends on the metric, and there great discrepancies~\cite{blau2018perception, blau20182018, gu2020image}. IQA models must learn to predict quantitative outputs as much correlated as possible with the MOS human ratings. We appreciate a huge perceptual quality difference between (b) and (d), however, neither PSNR nor SSIM reflect this.
    }
    \label{fig:teaser}
\end{figure*}

Image quality assessment (IQA) aims at using computational models to measure the perceptual quality of images, which are degraded during acquisition, generation, compression or post-processing operations~\cite{sheikh2006statisticalLIVE, zhai2020perceptual}.
Since one of the goals of the image processing is to improve the quality of the content to an acceptable level for the human viewers, IQA, as a ``evaluation technique'', plays a critical role in most image processing tasks such as image super-resolution, denoising, compression and enhancement~\cite{pipal, bhat2021ntire, gu2021ntire, yang2021ntire, blau20182018}. Although it is easy for human beings to distinguish perceptually better images, it has been proved to be difficult for algorithms \cite{tid2013,pipal, prashnani2018pieapp, tid2008}.\\

Recently, Generative Models~\cite{gan, esser2021taming, lugmayr2020srflow} have shown promising results for image enhancement and generation, producing realistic results to the human eye. For instance, perceptual image processing algorithms based on Generative Adversarial Networks (GANs)~\cite{gan, srgan2017, wang2018esrgan, cheon2018generative} have produced images with more realistic textures. 

However, these generated images show completely different characteristics and artifacts from traditional distortions (i.e. Gaussian Noise, Blur), for this reason, it has been noticed that the contradiction between the quantitative evaluation results and the real perceptual quality is increasing~\cite{blau20182018,blau2018perception,pipal}. Therefore, these methods have posed a great challenge for IQA methods to evaluate their visual quality. New IQA methods need to be proposed accordingly to evaluate new image processing algorithms, as this will also affect the development of such methods~\cite{blau2018perception,blau20182018 ,pipal, gu2020image, gu2021ntire}.

In this context, in order to generate acceptable images we have to accurately measure their perceptual quality, which can be performed via subjective and objective quality assessment~\cite{wang2011applications, cheon2017subjective, hu2020subjective, fang2020perceptual}. The subjective quality assessment is the most accurate method to measure the perceived quality, which is usually represented by mean opinion scores (MOS) from collected human subjective ratings. However, it is time-consuming and expensive.

Deep Convolutional Neural Networks (CNNs) can extract complex features from the images, and thus, they can provide a powerful IQA metric if there is enough data to train them. Moreover, these represent differentiable functions, allowing to plug them into adversarial training frameworks and optimize for quality directly~\cite{zhang2018unreasonable, ding2020image, prashnani2018pieapp, bosse2017deep, pipal}.

In general, we find two different IQA approaches: (i) \textit{Full-Reference (FR)}~\cite{gu2021ntire, sheikh2006statisticalLIVE, DeepQA2021ntire, cheon2021perceptual, tid2008} where an image without distortions is available besides the distorted image. (ii) \textit{No-Reference (NR)} (also known as Blind)~\cite{bosse2017deep, ma2017learning, bosse2017deep, brisque, hosu2020koniq} where only the distorted or generated image is available. Typically, Full-Reference approaches achieve better performance, however, Blind IQA (BIQA) represents the most realistic scenario and these approaches are more useful because of their feasibility. In Section~\ref{sec:rel_work} we present the \textit{state-of-the-art} of each case.

The \textbf{NTIRE 2022 Perceptual Image Quality Assessment Challenge}~\cite{gu2022ntire} seeks for novel solutions for Full-Reference and No-Reference IQA. In comparison with previous IQA benchmarks~\cite{tid2013, prashnani2018pieapp}, the training and testing datasets in this challenge include the outputs of GAN-based algorithms and the corresponding subjective scores, which provide more diversity and challenging scenarios. In this work we provide the following key contributions:
\begin{itemize}

\item In Section~\ref{sec:fullref} we introduce our conformer-based 4th place solution for Full-Reference IQA, as an alternative to transformer-based approaches like IQT~\cite{cheon2021perceptual} (winner of last year challenge).

\item In Section~\ref{sec:blind} we present our 3rd place solution for No-Reference IQA: Exploration of semi-supervised noisy student learning to distill knowledge from FR models into blind noisy student models.

\item Comparison with the NTIRE 2021 IQA Challenge~\cite{gu2021ntire} methods and extensive ablation studies.

\end{itemize}

\section{Related Work}
\label{sec:rel_work}

\paragraph{Image Quality Assessment.}

CNNs have shown their effectiveness in a wide range of computer vision and image processing tasks, such as super-resolution, denoising and deblurring~\cite{nah2021ntire, yang2021ntire, timofte2017ntire}. Generative models~\cite{esser2021taming, lugmayr2020srflow,gan}, and in particular, GAN-based~\cite{gan, srgan2017} approaches produce typically more pleasant results to human eyes than the CNNs that do not use adversarial loss.
The goal of the developing IQA methods is to accurately predict the perceived quality (by human viewers) of the generated images. However, traditional IQA methods struggle to evaluate these new approaches, and there are contradictions between the perceptual quantitative results and the qualitative results.
We can classify IQA methods depending on:
\begin{itemize}

    \item Input data: (i) Full-Reference (FR)~\cite{gu2021ntire, sheikh2006statisticalLIVE, DeepQA2021ntire, cheon2021perceptual, tid2008} where a reference image without distortions is available besides the distorted image. (ii) No-Reference (NR)~\cite{bosse2017deep, ma2017learning, bosse2017deep, brisque, hosu2020koniq} where only the distorted or generated image is available.
    
    \item Training: (i) Traditional methods do not require training. (ii) Learnable methods (typically CNN-based).
\end{itemize}

\paragraph{Full-Reference IQA}

The FR methods focus more on visual similarity or dissimilarity between two images (typically the original or reference image, and the generated one). The most representative IQA FR metrics are the PSNR, which is related to the MSE between both images, and the SSIM proposed by Wang \textit{et al.}~\cite{wang2004image}. 
These traditional methods have the advantage of convenience for optimization; however, they poorly predict humans perceived visual quality, especially for evaluating fine textures and details in the images~\cite{1609.04802}.
Since that, various FR metrics have been developed to take into account various aspects of human quality perception, e.g., information-theoretic criterion \cite{sheikh2005information} or structural similarity \cite{wang2003multiscale, zhang2011fsim}.

Note that the ultimate goal of image enhancement networks is to generate visually pleasant images for humans and have a high MOS, which is not always strictly correlated to these traditional metrics.
Recently, learned CNN-based IQA methods have been actively studied and provide the most promising \textit{state-of-the-art} results~\cite{zhang2018unreasonable, bosse2017deep, prashnani2018pieapp, ding2021comparison, hosu2020koniq, gu2021ntire}.
Zhang~\textit{et al.} proposed a learned perceptual image patch similarity (LPIPS) metric~\cite{zhang2018unreasonable}, which shows that trained deep features that are optimized by the ${l}_2$ distance between distorted and reference images are effective for IQA compared to the conventional IQA methods.

Among the most competitive approaches in the NTIRE 2021 IQA Challenge~\cite{gu2021ntire} we can find: ASNA~\cite{asna2021ntire} proposed a CNN equipped with spatial and channel-wise attention mechanisms, and Siamese-like network architecture. IQMA~\cite{IQMA2021ntire} proposed a bilateral-branch multi-scale image quality estimation network, using Feature Pyramid Network (FPN)-like architecture to extract multi-scale features and predict the the quality score of the image at multiple scales.

Cheon~\textit{et al.} introduced an image quality transformer IQT~\cite{cheon2021perceptual} that successfully applies a transformer architecture to a perceptual full-reference IQA task. This method combines a CNN backbone as a feature extractor, with a Transformer~\cite{vaswani2017attention} encoder-decoder to compare a reference and distorted images, and predict the quality score.

\paragraph{Blind IQA}

The No-Reference (NR) or Blind methods~\cite{hou2014blind, xin2002blind} are useful because of its feasibility, they can be plugged-in in adversarial training frameworks and be used for optimizing perceptual quality directly. However, the absence of a reference image makes it challenging to predict image quality accurately compared to the FR methods.

Bosse~\textit{et al.}~\cite{bosse2017deep} studies the performance of deep neural networks for no-reference and full-reference image quality assessment.
Mittal~\textit{et al.}~\cite{brisque} explores blind IQA in the spatial domain.
Zhang~\textit{et al.}~\cite{zhang2021uncertainty} proposes a model and a training approach to deal with realistic and synthetic distortions and improve the generalization capabilities.

The NTIRE 2022 IQA Challenge introduced this year a track for Blind image quality assessment (BIQA). 

\paragraph{Evaluation}

IQA methods should present the following two desired characteristics: (i) high Pearson linear correlation coefficient (PLCC) between the scores produced by the proposed method and the ground-truth MOS, which indicates the linear relationship between them, (ii) high Spearman rank order correlation coefficient (SRCC), which shows the monotonicity of relationship between the proposed method and the ground-truth MOS.
Both metrics separately, and the sum of both as a "Main Score'', serve as evaluation metric to compare the performance of IQA methods~\cite{gu2021ntire, sheikh2006statisticalLIVE, pipal}. The Kendall Rank-order Correlation Coefficient (KRCC) is also used to estimate the monotonicity and consistency of the quality prediction~\cite{gu2020image}.

\paragraph{Datasets} 

TID2013 \cite{tid2013}, LIVE~\cite{sheikh2006statisticalLIVE} and PIPAL \cite{pipal, gu2020image} provide images with their corresponding reference images and MOS to train models in a supervised manner. We compare these datasets in Table~\ref{tab:datasets}.
The NTIRE 2022 IQA Challenge~\cite{gu2022ntire} uses the PIPAL~\cite{pipal} dataset, which takes a step forward in benchmarking perceptual IQA by incorporating the perceptual quality of images obtained by perceptual-oriented algorithms (i.e. GANs), missing in previous datasets. The PIPAL~\cite{pipal} as our training set, contains 200 reference images, 23k distorted images and their respective human judgements. 
To ensure that the models can generalize properly, the challenge has an extended dataset of PIPAL for validation and testing. This dataset contains 3300 distorted images (1650 for training and testing respectively) for 50 reference images, and all of them are the outputs of perceptual-oriented algorithms. It collects 753k human judgements to assign subjective scores for the extended images, ensuring the objectivity of the testing data. Participants do not have access to the ground-truth for validation or test, results are submitted using a public website.

\begin{table}[ht]
	\centering
	\resizebox{\linewidth}{!}{
	\begin{tabular}{lcccccc}
		\toprule
		Database    & \# Ref.  &  \# Dist.   &   Dist.~Type    &   \# Dist.~Type &  \# Rating \\ 
		\midrule
		LIVE~\cite{sheikh2006statisticalLIVE} &	29 & 779 & trad. & 5 & 25k \\
		TID2013~\cite{tid2013}	& 25 & 3k & trad. & 25  & 524k\\ 
		PIPAL~\cite{pipal}	&250 &	29k	& trad.+alg.  & 40  & 1.13m	\\ 
		\bottomrule
	\end{tabular}}
	\caption{Comparison of IQA datasets for performance evaluation. The NTIRE Challenge Dataset, PIPAL~\cite{pipal} presents the highest and more various number of distorsions and human ratings.}
	\label{tab:datasets}
\end{table}

\section{IQA Conformer Network}
\label{sec:fullref}

We propose an alternative architecture to IQT~\cite{cheon2021perceptual} by replacing the Transformer encoder-decoder~\cite{vaswani2017attention} by a Conformer architecture~\cite{gulati2020conformer, burchi2021efficient}, which uses convolution and attention operations to model local and global dependencies. 

We use a Inception-ResNet-v2~\cite{szegedy2017inception} network pre-trained on ImageNet to extract feature maps from the reference and distorted image. The network weights are kept frozen and a Conformer~\cite{gulati2020conformer} encoder-decoder is trained to regress MOS using the MSE loss. As done by Cheon~\textit{et al.}~\cite{cheon2021perceptual}, we concatenate the feature maps from the following blocks: \texttt{mixed5b, block35$\_$2, block35$\_$4, block35$\_$6, block35$\_$8 and block35$\_$10}. We do this for the reference and distorted images generating $f_{ref}$ and $f_{dist}$, respectively. In order to obtain difference information between reference and distorted images, a difference feature map, $f_{diff}=f_{ref}-f_{dist}$ is also used. 

\begin{figure}[!ht]
    \centering
    \setlength{\tabcolsep}{2.0pt}
    \begin{tabular}{c}
    \includegraphics[width=\linewidth]{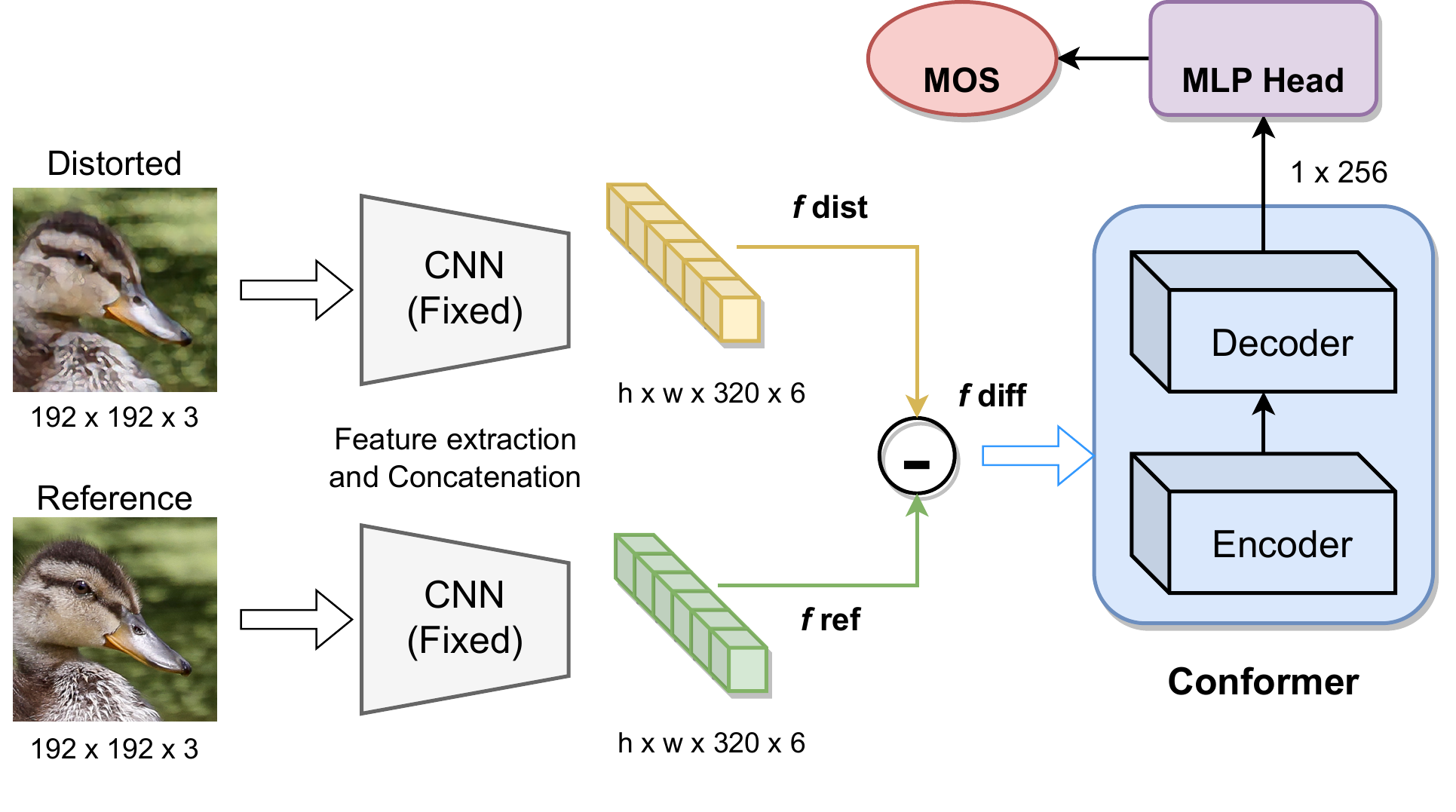} \tabularnewline
    \end{tabular}
    \caption{FR IQA Conformer setup inspired by IQT~\cite{cheon2021perceptual}.}
    \label{fig:iqc}
\end{figure}

\begin{figure*}[!ht]
    \centering
    \setlength{\tabcolsep}{2.0pt}
    \begin{tabular}{cccc}
    \includegraphics[width=0.2\linewidth]{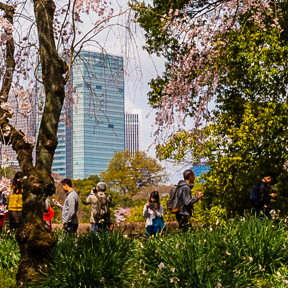} &
    \includegraphics[width=0.2\linewidth]{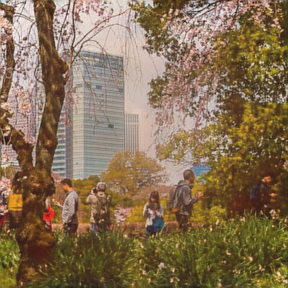} &
    \includegraphics[width=0.2\linewidth]{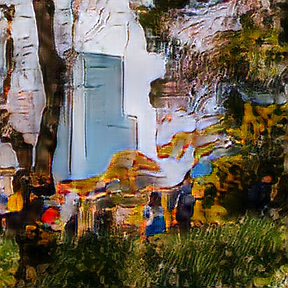} &
    \includegraphics[width=0.2\linewidth]{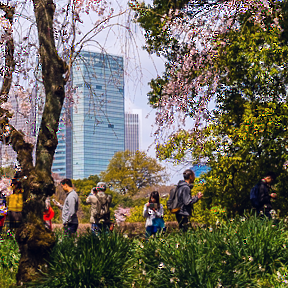} \tabularnewline
    Reference & PSNR 27.972 - SSIM 0.735 & PSNR 28.512 - SSIM 0.302 & PSNR 31.496 - SSIM 0.861 \tabularnewline
     & Ours MOS 1368.53  & Ours MOS 1353.89 & Ours MOS 1564.38
    \tabularnewline
    \includegraphics[width=0.2\linewidth]{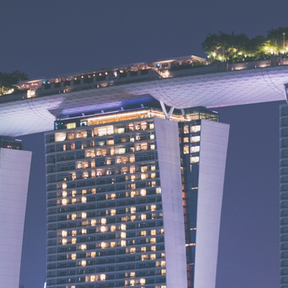} &
    \includegraphics[width=0.2\linewidth]{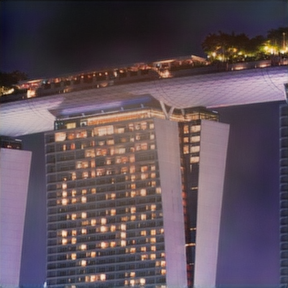} &
    \includegraphics[width=0.2\linewidth]{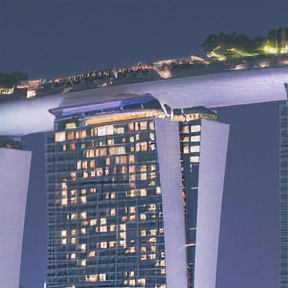} &
    \includegraphics[width=0.2\linewidth]{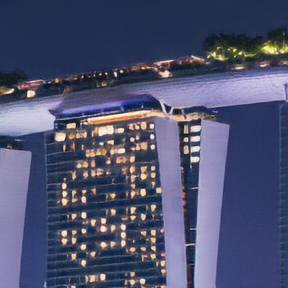} \tabularnewline
    Reference & PSNR 27.968 - SSIM 0.867 & PSNR 32.316 - SSIM 0.815 & PSNR 27.836 - SSIM 0.842 \tabularnewline
     & Ours MOS 1283.033  & Ours MOS 1525.727 & Ours MOS 1534.356 \tabularnewline
    \end{tabular}
    \caption{Example images from the test set of the NTIRE 2022 challenge~\cite{gu2022ntire}. For each distorted image we provide: PSNR, SSIM and our predicted MOS. Our model is the most correlated quantitative metric to the real human MOS subjective ratings.
    }
    \label{fig:samples}
\end{figure*}

\begin{figure*}[!ht]
    \centering
    \setlength{\tabcolsep}{2.0pt}
    \begin{tabular}{cc}
    \includegraphics[width=0.48\linewidth]{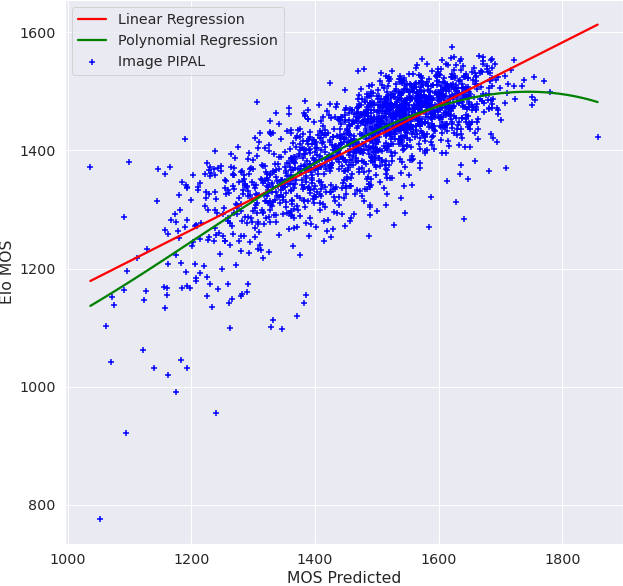}&
    \includegraphics[width=0.48\linewidth]{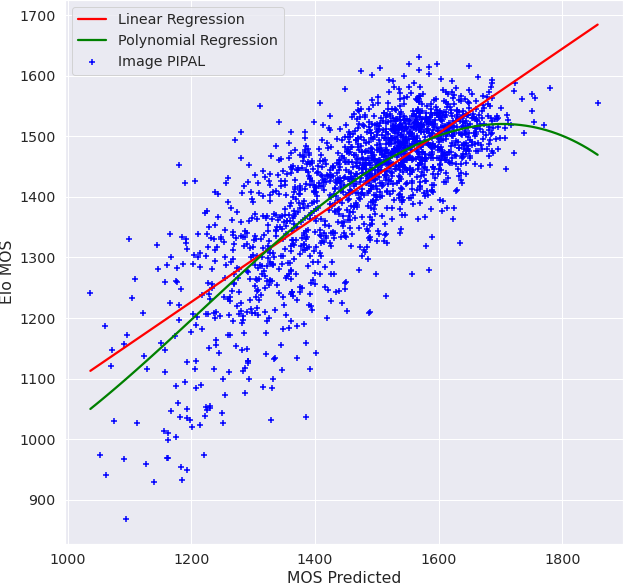}
    \tabularnewline
    PIPAL~\cite{pipal} Full-Reference & PIPAL~\cite{pipal} No-Reference
    \tabularnewline
    \end{tabular}
    \vspace{-0.2cm}
    \caption{Predicted MOS scores against Elo MOS subjective scores on the validation set of PIPAL~\cite{pipal} NTIRE 2022 IQA Challenge~\cite{gu2022ntire}.}
    \label{fig:mos-scatter}
    \vspace{-0.3cm}
\end{figure*}

Concatenated feature maps are then projected using a point-wise convolution but not flattened to preserve spatial information. We used a single Conformer block~\cite{gulati2020conformer} for both encoder and decoder. The model hyper-parameters are set as follows: $L=1$, $D=128$, $H=4$, $D_{feat}=512$, and $D_{head}=128$. The input image size of the backbone model is set to (192 × 192 × 3) which generates feature maps of size $21$ x $21$. \textbf{IQA Conformer} has 2,831,841 total parameters, we illustrate the pipeline in Figure~\ref{fig:iqc}. Note that we only use the first feature maps from the CNN, not the whole network, therefore the number of parameters is substantially smaller.
%
In Table~\ref{tab:benchmark2021} we compare our method with the \textit{state-of-the-art} on the NTIRE 2021 and 2022 IQA Challenges~\cite{gu2021ntire, gu2022ntire}. For a fair comparison, all the models were trained using the same PIPAL training dataset~\cite{pipal, gu2021ntire}. We use RADN~\cite{shi2021radn} and ASNA~\cite{ayyoubzadeh2021asna} public available pre-trained weights and code. Our proposed solution allows to reach better PLCC and SRCC at inference than IQT~\cite{cheon2021perceptual} under the same setup (see Table~\ref{tab:benchmark2021}). Note that to the best of our knowledge, there is no public code or models for reproducing IQT~\cite{cheon2021perceptual} results, therefore we report results of our best implementation following the original paper.
%
We also compare our results with other top performing teams at the NTIRE 2022 IQA FR Challenge~\cite{gu2022ntire} in Table~\ref{tab:lbfinal_fullref}, where our IQA Conformer was ranked 4th. We show qualitative samples and analysis in Figures~\ref{fig:samples} and~\ref{fig:mos-scatter}.

\begin{table}[ht]
	\centering
	\begin{tabular}{lccc}
		\toprule
		Team          & Main Score~$\uparrow$ &  PLCC & SRCC \\
		\midrule
        THU1919Group & 1.651 & 0.828 & 0.822 \\
        Netease OPDAI & 1.642 & 0.827 & 0.815 \\
        KS & 1.640 & 0.823 & 0.817 \\
        \textbf{Ours} & 1.541 & 0.775 & 0.766 \\
        Yahaha! & 1.538 & 0.772 & 0.765 \\
        debut\_kele & 1.501 & 0.763 & 0.737 \\
        Pico Zen & 1.450 & 0.738 & 0.713 \\ 
        Team Horizon & 1.403 & 0.703 & 0.701 \\
        \bottomrule
	\end{tabular}
	\caption{Performance comparison of the top teams on the testing dataset of the NTIRE 2022 Full-Reference IQA Challenge.}
	\label{tab:lbfinal_fullref}
\end{table}

\paragraph{Implementation details}
The model was trained using only the NTIRE 2022 PIPAL training dataset~\cite{pipal}. Adam optimizer by setting $\beta_1= 0.9$, $\beta_2= 0.999$. We set minibatch size as 16. The learning rate was set to $10^{-4}$ and the model trained for 30 epochs (43479 gradient steps). Last 10 epoch checkpoints were averaged using SWA~\cite{izmailov2018averaging}.

\paragraph{Inference} 
During inference, we use \textit{enhanced prediction}~\cite{timofte2016seven} (a.k.a Test-Time Augmentations). The prediction for an input image is enhanced by averaging the predictions on a set of transformed images derived from it. We use 10 crops (2 flips of 4 image corners + center crop) for reference and distorted images.

\paragraph{Ensembles and fusion strategies} 
As shown in Table~\ref{tab:benchmark2021} an ensemble of our model, RADN~\cite{shi2021radn} and ASNA~\cite{ayyoubzadeh2021asna}, improves notably the performance (+0.4 boost in main score).

\subsection{Cross Database Evaluations}

\begin{figure}[!hb]
    \centering
    \includegraphics[width=\linewidth]{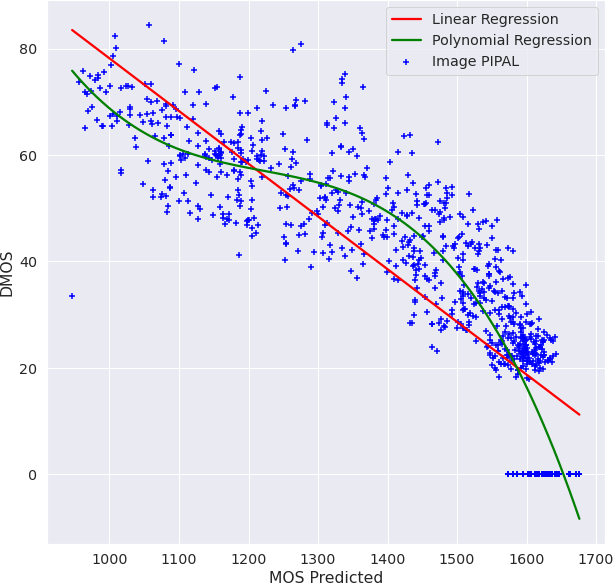}
    \caption{Ground-truth DMOS LIVE~\cite{sheikh2006statisticalLIVE} against the predicted MOS. Our predictions have $|SRCC| = 0.92 $, which indicates they are very correlated with the real qualitative ratings.}
    \label{fig:mos-live}
\end{figure}

IQA methods tend to overfitting, they commonly struggle to generalize to data distributions different from the one they were trained with.
To validate the generalization capabilities of our approach, we use our FR IQA Conformer trained on PIPAL~\cite{pipal} and we conduct the cross-dataset evaluation on two other benchmarks: TID2013~\cite{tid2013} and LIVE~\cite{sheikh2006statisticalLIVE} (using the full datasets).
%
As shown in Table~\ref{tab:cross_eval}, our model generalizes better than NTIRE 2021~\cite{gu2021ntire} top methods: ASNA~\cite{asna2021ntire}, RADN~\cite{RADN2021ntire} and IQT~\cite{cheon2021perceptual}.
It also achieves competitive results in comparison with other learnt methods like PieAPP~\cite{prashnani2018pieapp}, WaDIQaM~\cite{bosse2017deep} or LPIPS~\cite{zhang2018unreasonable}, which are trained on the specific datasets.
Figure~\ref{fig:mos-live} shows a qualitative analysis of the predictions for LIVE~\cite{sheikh2006statisticalLIVE}.

\begin{table}[!hb]
	\centering
	\begin{tabular}{lcccc}
		\toprule
		\multirow{2}{*}{Method}&\multicolumn{2}{c}{LIVE~\cite{sheikh2006statisticalLIVE}}&\multicolumn{2}{c}{TID2013~\cite{tid2013}}\\ 
		&SRCC&KRCC&SRCC&KRCC\\ \hline
		PSNR~\cite{Hore2010}	&	0.873	&	0.680 &	0.687	&	0.496	\\
		SSIM \cite{wang2004image}	&	0.948	&	0.796 &	0.727	&	0.545	\\
		MS-SSIM \cite{wang2003multiscale}	&	0.951	&	0.805	&	0.786	&	0.605	\\
		VIF \cite{sheikh2006statisticalLIVE} &		0.964	&	0.828	&	0.677	&	0.518	\\
		NLPD \cite{laparra2016perceptual}	&	0.937	&	0.778	&	0.800	&	0.625	\\
		GMSD \cite{xue2014gradient}	&	0.960	&	0.827	&	0.804	&	0.634	\\
		\midrule
		WaDIQaM \cite{bosse2017deep}	&	0.947	&	0.791	&	0.831	&	0.631	\\
		PieAPP \cite{prashnani2018pieapp}	&	0.919	&	0.750	& 0.876	&	0.683	\\
		LPIPS \cite{zhang2018unreasonable}	&	0.932	&	0.765	&	0.670	&	0.497	\\
		DISTS \cite{ding2021comparison}	&	0.954	&	0.811	&	0.830	&	0.639	\\
		SWDN~\cite{gu2020image}	&	-	&	-	&	0.819	&	0.634	\\
		\textcolor{blue}{ASNA}~\cite{asna2021ntire} & 0.92 & - & 0.73 & - \\
		\textcolor{blue}{RADN}~\cite{RADN2021ntire} & 0.905 & - & 0.747 & - \\
		\textcolor{blue}{IQT-C}~\cite{cheon2021perceptual} &  0.917	& 0.737 & 0.804 & 0.607 \\ 
		\midrule
		Ours  &  0.921	& 0.752 & 0.82 & 0.630 \\
		\bottomrule
	\end{tabular}
	\caption{Performance comparison on LIVE~\cite{sheikh2006statisticalLIVE} and TID2013~\cite{tid2013}. Some results are borrowed from \cite{ding2020image, gu2020image}. We separate traditional and learnt methods, and we highlight in \textcolor{blue}{blue} the NTIRE 2021~\cite{gu2021ntire} methods trained on PIPAL~\cite{pipal}.}
	\label{tab:cross_eval}
\end{table}

\subsection{Ablation Study}

In Table~\ref{tab:benchmark2021} we compare the performance of Transformer~\cite{vaswani2017attention, cheon2021perceptual} and Conformer~\cite{gulati2020conformer} models, both using the same backbone (Inception-ResNet-v2~\cite{szegedy2017inception}) and training setup.
We also explore the effect of different backbone architectures for feature extraction like ConvNext~\cite{liu2022convnext} (SOTA in image classification) and VGG~\cite{vgg} (common backbone in IQA). We find Inception-ResNet-v2~\cite{szegedy2017inception} features to be the best representation.
These approaches are very sensitive to the backbone selection, and more specifically, to the feature block selection (as introduced by EGB~\cite{EGB2021ntire}).

\begin{table*}[!ht]
	\centering
	\begin{tabular}{@{}lcccccccc@{}}
		\toprule
		\multirow{2}{*}{Method} & \multicolumn{2}{c}{Validation 2021} & \multicolumn{2}{c}{Testing 2021} & \multicolumn{2}{c}{Validation 2022} & \multicolumn{2}{c}{Testing 2022} \\ 
		                                    &   PLCC    &   SRCC    & PLCC      &   SRCC &   PLCC    &   SRCC    & PLCC      &   SRCC  \\
		\midrule
        PSNR~\cite{Hore2010}	                            &   0.292   &   0.255   &   0.277	&	0.249 & 0.269	&	0.234	&	0.277	&	0.249	\\
        \rowcolor{Gray}
        SSIM~\cite{wang2004image}	        &   0.398   &   0.340   &   0.394	&	0.361 & 0.377	&	0.319	&	0.391	&	0.361	\\
        VSI~\cite{zhang2014vsi}	            &   0.516   &   0.450   &   0.517	&	0.458 &  0.493	&	0.411	&	0.517	&	0.458	\\
        \rowcolor{Gray}
        NQM~\cite{damera2000image}	        &   0.416   &   0.346   &   0.395	&	0.364 & 0.364	&	0.302	&	0.395	&	0.364	\\
        UQI~\cite{wang2002universal}	    &   0.548   &   0.486   &   0.450	&	0.420 & 0.505	&	0.461	&	0.450	&	0.420	\\
        \rowcolor{Gray}
        GSM~\cite{liu2012image}	            &   0.469   &   0.418   &   0.465	&	0.409 & 0.450	&	0.379	&	0.465	&	0.409	\\
        RFSIM~\cite{zhang2010rfsim}	        &   0.304   &   0.266   &   0.328	&	0.304 & 0.285	&	0.254	&	0.328	&	0.304	\\
        \rowcolor{Gray}
        SRSIM~\cite{zhang2012sr}	        &   0.654   &   0.566   &   0.636	&	0.573 & 0.626	&	0.529	&	0.636	&	0.573	\\
        PI~\cite{blau20182018}	            &   0.166   &   0.169   &   0.145	&	0.104 & 0.134	&	0.064	&	0.145	&	0.104	\\
        \rowcolor{Gray}
        LPIPS-VGG~\cite{zhang2018unreasonable}	&	0.647	&	0.591	&	0.633	&	0.595 & 0.611	&	0.551	&	0.633	&	0.595	\\
        DISTS~\cite{ding2021comparison}	    &	0.686	&	0.674	&	0.687	&	0.655 & 0.634	&	0.608	&	0.687	&	0.655	\\
        \rowcolor{Gray}
        \textcolor{blue}{EGB}~\cite{EGB2021ntire} &  0.775    &  0.776     &   0.677    &   0.700 & 0.746 & 0.723 & 0.700 & 0.677 \\ 
        \textcolor{blue}{ASNA}~\cite{ayyoubzadeh2021asna} &   0.820    &   0.830    &   0.750    &   0.710 & 0.796 & 0.765 & 0.752 & 0.719\\ 
        \rowcolor{Gray}
        \textcolor{blue}{RADN}~\cite{shi2021radn} & 0.866 & 0.865 & 0.771 & 0.777 & 0.789 & 0.777 & 0.753 & 0.757\\ 
        \textcolor{blue}{IQT (2021 Winner)}~\cite{cheon2021perceptual} & 0.876     &  0.865    & 0.790     & 0.799 &  0.840	&	0.820	&	\textbf{0.799}	&	\textbf{0.790} \\ 
        \midrule
        Ours IQA Transformer & & & & & 0.790 & 0.765 & 0.757 & 0.751 \\
        Ours IQA Conformer A & & & & & 0.804 & 0.790 & \textbf{0.775} & \textbf{0.766} \\
        Ours IQA Conformer B & & & & & 0.740 & 0.740  & 0.730 & 0.730 \\
        Ours IQA Conformer C & & & & & 0.790 & 0.770 & 0.754 & 0.740 \\
        Ensemble & & & & & & & \textbf{0.787} & \textbf{0.793} \\
        \bottomrule
	\end{tabular}
	\caption{Performance comparison of IQA methods on the PIPAL NTIRE 2021 and 2022 Full-Reference benchmark~\cite{pipal, gu2021ntire, gu2022ntire}. We highlight in \textcolor{blue}{blue} the top performing methods on the NTIRE 2021 IQ Challenge~\cite{gu2021ntire}. The different IQA Conformer versions correspond to different backbones: (A) Inception-ResNet-v2~\cite{szegedy2017inception}, (B) ConvNext~\cite{liu2022convnext}, (C) VGG19~\cite{vgg}.
	The ensemble method is: Ours + RADN~\cite{shi2021radn} + ASNA~\cite{ayyoubzadeh2021asna}. Ours IQA "Transformer'' is our own implementation of IQT~\cite{cheon2021perceptual}, since, to the best of our knowledge, there is not public code available.
	}
	\label{tab:benchmark2021}
\end{table*}

\begin{figure*}[!ht]
    \centering
    \includegraphics[width=0.62\linewidth]{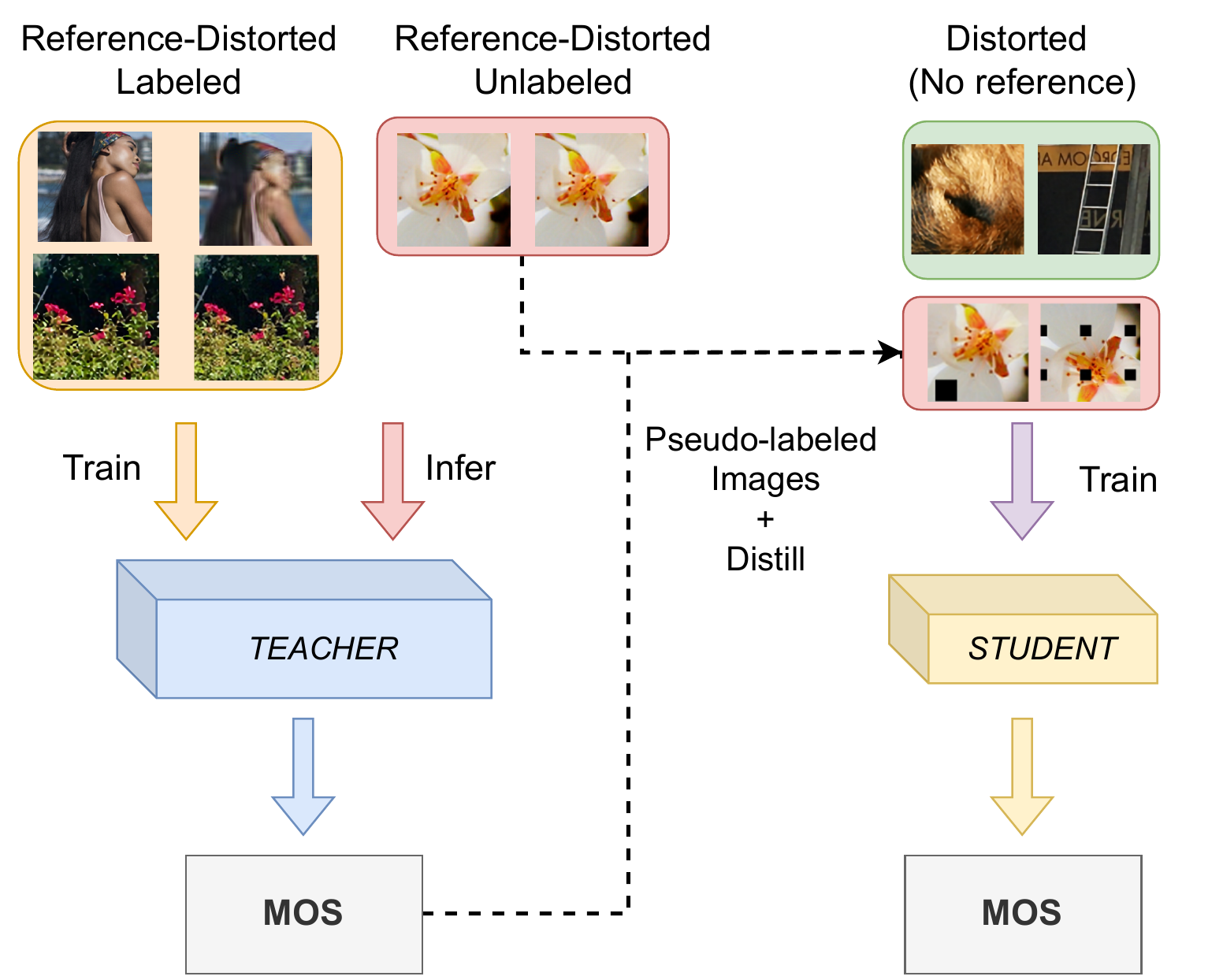}
    \caption{Full-Reference Teacher and Blind Noisy Student. Unlabeled samples are annotated using pseudo-labels inferred using the teacher.}
    \label{fig:iq-noisy-student}
\end{figure*}

\section{Blind Noisy Student}
\label{sec:blind}

\begin{figure}[!ht]
    \centering
    \includegraphics[width=\linewidth]{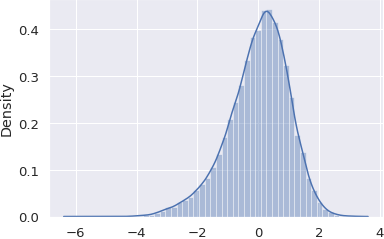}
    \caption{PIPAL~\cite{pipal} training MOS standardized  distribution with $\mu = 1448.96$ and  $\sigma = 121.53$. We consider perceptual high-quality with MOS~$> 2$, and low-quality MOS~$<-2$.
    }
    \label{fig:mos-dist}
\end{figure}

A simple CNN backbone $\Phi$ takes as input a distorted image $x$ and aims to minimize the MOS $y$ (see Figure~\ref{fig:mos-dist}) using the following loss function from~\cite{ayyoubzadeh2021asna}, where $\Phi(x) = \hat{y}$.

\begin{equation}
    \mathcal{L} = MSE(y, \hat{y}) + (1 - Pearson(y, \hat{y}))
\label{eq:loss}
\end{equation}

\paragraph{Initial setup}
We train EfficientNet B0~\cite{tan2019efficientnet} (pre-trained on ImageNet) to perform this task. Using a 90/10 validation split (i.e. validating on roughly 1000 images locally), we achieved $1.02$ on the development phase using a single model. We use as augmentations in all our experiments the following pipeline: random horizontal and vertical flips, random rotations of 90/180/270 degrees, and finally take a random crop of size $224$ x $224$.
We find the main performance limitation to be overfitting due to the small dataset: 23200 images, yet only 200 reference images~\cite{pipal}.
We select EffNet B0~\cite{tan2019efficientnet} as backbone as it is \textit{stat-of-the-art} in Image classification and has only 4 million parameters.

\begin{figure}[!ht]
    \centering
    \includegraphics[width=0.85\linewidth]{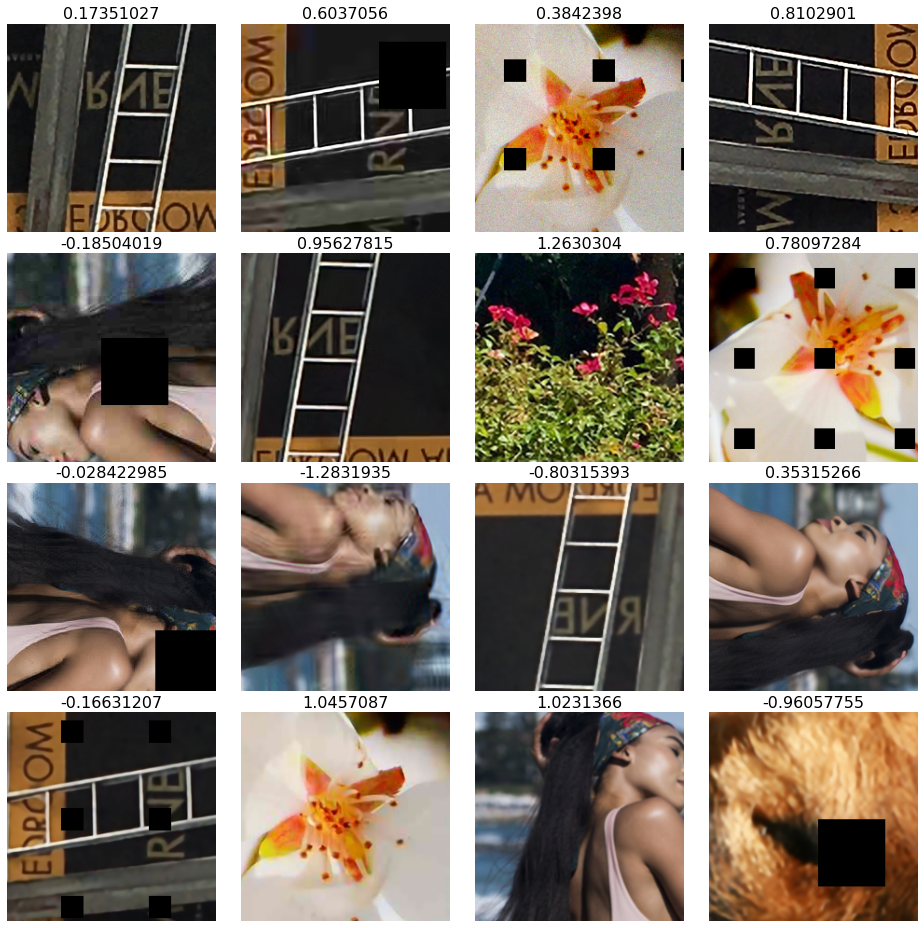}
    \caption{Example of ''Extra" augmentations~\cite{devries2017cutout} for the noisy student training. We show standardized  MOS above each image. }
    \label{fig:augs}
\end{figure}

\paragraph{Noisy student} 
a semi-supervised learning approach that extends the idea of self-training and distillation, and has achieved \textit{state-of-the-art} results on Image Classification~\cite{xie2020self}.
We distinguish a Full-Reference \textbf{teacher} model, and a blind \textbf{student} model trained only with distorted images. This method allows to increase the amount of training distorted images, and to transfer ”dark” knowledge~\cite{xie2020self} from FR models and ensembles, into simple NR models. The process is as follows:

\begin{enumerate}
    \item Train the FR teacher using the training dataset~\cite{pipal, gu2021ntire} consisting on 23k reference-distorted pairs.
    \item Teacher infers on \textbf{unlabeled} reference-distorted pairs, and annotate the images. These MOS annotations are noisy, we refer to them as \textit{pseudo-labels}.
    \item Add the pseudo-labeled distorted images to the original training set: approximately 2k new images.
    \item Train a student model for NR IQA, which takes as input only the distorted images using the extended dataset (original + pseudo-labels) and extra augmentations to the initial setup: (i) CutOut~\cite{devries2017cutout} as further regularization to ensure the model learns useful features without looking to the entire image. (ii) Small perturbations on the Saturation, Brightness and Contrast. We show some examples in Figure~\ref{fig:augs}.
\end{enumerate}

We illustrate this process in Figure~\ref{fig:iq-noisy-student}. Using this approach we can distill knowledge from the FR models into the NR models. We show the benefits of this approach and augmentations in Table~\ref{tab:ablation_blind_students}.
We obtain the unlabeled samples from two different ways: (i) using the unlabeled data provided at the challenge and PIPAL~\cite{pipal,gu2021ntire}, (ii) augmenting the reference images using traditional methods and GAN-based methods like SRGAN~\cite{srgan2017, wang2018esrgan} to upscale the images and resize back to the original resolution. We do not use other datasets for training.

Therefore, our single model EffNet B0~\cite{tan2019efficientnet} has only 4 million parameters, in comparison with other well-known architectures used for this task such as VGG~\cite{simonyan2014very} (15 million parameters) or ResNet 50~\cite{he2016deep} (24 million parameters). In our experiments, deep models tend to overfitting quickly and did not perform great. In Table~\ref{tab:lbfinal_noref} we show our performance in comparison with other top teams.

\begin{table}[!ht]
    \centering
    \resizebox{\columnwidth}{!}{%
    \begin{tabular}{lccccc}
         \toprule
         Method & \# Extra & Augs. & Score~$\uparrow$ & PLCC & SRCC \\
         \midrule
         EffNet B0~\cite{tan2019efficientnet} & 0 & \xmark  & 0.84  & 0.42 & 0.42 \\
         EffNet B0~\cite{tan2019efficientnet} & 0 & \cmark & 1.02 & 0.51  & 0.51 \\
         EffNet B0~\cite{tan2019efficientnet} & 1.6k & \cmark & 1.42 & 0.73 & 0.70 \\
         VGG 19~\cite{simonyan2014very} & 1.6k & \cmark & 1.25 & 0.63 & 0.61 \\
         ResNet50~\cite{he2016deep} & 1.6k & \cmark & 1.37 & 0.70 & 0.67 \\
         EffNet B0~\cite{tan2019efficientnet} & 2k   & \cmark & 1.48 & 0.75 & 0.73 \\
         EffNet B0~\cite{tan2019efficientnet}~+~TA & 2k   & \cmark & 1.49 & 0.76 & 0.73 \\
         \bottomrule
    \end{tabular}
    }
    \caption{Ablation study of our NR models. We indicate the number of extra pseudo-labeled samples added to the original training dataset, the use of ''extra" augmentations, and the scores for each model in the NTIRE 2022 IQA Challenge test set. TA~\cite{timofte2016seven} indicates Test-time Augmentations (i.e. average of 4 random crops).}
    \label{tab:ablation_blind_students}
\end{table}

\begin{table}[!ht]
	\centering
	\begin{tabular}{lccc}
		\toprule
		Team          & Main Score~$\uparrow$ &  PLCC & SRCC \\
		\midrule
        THU\_IIGROUP   &  1.444  &  0.740  &  0.704 \\
        DTIQA          &  1.437  &  0.737  &  0.700 \\
        \textbf{Ours}  &  1.422  &  0.725  &  0.697 \\
        KS             &  1.407  &  0.726  &  0.681 \\
        NetEase OPDAI  &  1.390  &  0.720  &  0.671 \\
        Minsu Kwon     &  1.183  &  0.607  &  0.576 \\
        NTU607QCO-IQA &  1.112  &  0.585  &  0.527 \\
        \bottomrule
	\end{tabular}
	\caption{Performance comparison of the top teams on the testing dataset of the NTIRE 2022 No-Reference IQA Challenge Main score is calculated as the sum of PLCC and SRCC.}
	\label{tab:lbfinal_noref}
\end{table}

\begin{table*}[!ht]
	\centering
	\begin{tabular}{@{}lcccccccc@{}}
		\toprule
		\multirow{2}{*}{Method} & \multicolumn{3}{c}{Validation 2022} & \multicolumn{3}{c}{Testing 2022} \\ 
		& Main Score~$\uparrow$ & PLCC & SRCC & Main Score~$\uparrow$ & PLCC & SRCC \\
		\midrule
        Brisque~\cite{brisque}	&	0.075	&	0.059	&	0.015	&	0.184	&	0.097	&	0.087	\\
        \rowcolor{Gray}
        NIQE~\cite{mittal2012making}	&	0.120	&	0.115	&	0.005	&	0.142	&	0.112	&	0.030	\\
        PI~\cite{blau20182018}	&	0.213	&	0.133	&	0.079	&	0.276	&	0.153	&	0.123	\\
        \rowcolor{Gray}
        Ma~\cite{ma2017learning} 	&	0.261	&	0.131	&	0.129	&	0.398	&	0.224	&	0.174	\\
        PSNR~\cite{Hore2010}	&	0.533	&	0.284	&	0.250	&	0.572	&	0.303	&	0.269	\\
        \rowcolor{Gray}
        SSIM~\cite{wang2004image}	&	0.718	&	0.386	&	0.332	&	0.785	&	0.407	&	0.377	\\
        FSIM~\cite{zhang2011fsim}	&	1.048	&	0.575	&	0.473	&	1.138	&	0.610	&	0.528	\\
        \rowcolor{Gray}
        LPIPS-AlexNet~\cite{zhang2018unreasonable}	&	1.197	&	0.616	&	0.581	&	1.176	&	0.592	&	0.584	\\
        \midrule
        Ours	&	\textbf{1.410}	&	0.710	&	0.700	&	\textbf{1.490}	&	0.752	&	0.733	\\
        \bottomrule
	\end{tabular}
	\vspace{-0.1cm}
	\caption{Performance comparison of IQA methods on the PIPAL NTIRE 2022 No-Reference benchmark~\cite{pipal, gu2022ntire}. Our method outperforms traditional and learnt methods by large margin. See also Table~\ref{tab:lbfinal_noref}, where we compare our method with other outstanding approaches.}
	\label{tab:benchmark-nr2022}
\end{table*}

\paragraph{Implementation details} 
We train each NR model to convergence, approximately 20 epochs. We use Adam optimizer with default parameters and learning rate $0.0001$. We set batch size to 32. The learning rate is reduced by factor 0.5 on plateaus. The loss function is presented in Equation~\ref{eq:loss}. We use a Tesla P100 GPU to run all our experiments. This model allows to predict the quality of the test set (1650 images) in just 8s (143ms/step) on a single GPU.

This inference time of approximately 0.22ms per image for a NR single model represents a beneficial approach for adversarial networks~\cite{gan}, where this model can be plugged-in as discriminator or differentiable loss function for direct perceptual quality optimization.

\section{Conclusion}

In this paper, we propose a method for IQA knowledge distillation from Full-Reference (FR) teacher models into Referenceless student models. 
First, we explore different IQA Full-Reference models, including transformer-based approaches. Next, we apply a semi-supervised noisy student approach: we annotate unlabeled reference-distorted image pairs using the FR model, we expand the original training set of distorted images using such pseudo-labeled data, and we finally train a Blind noisy student model.

Our methods achieved competitive performance on the latest PIPAL dataset, which contains new algorithm-based distorted images, and our predictions are well correlated with subjective human mean opinion scores of the images. Our methods achieved the 4th and 3rd place at the NTIRE 2022 Perceptual Image Quality Assessment Challenge for Full-reference and No-Reference respectively. Moreover, our approach can successfully generalize to other datasets like TID2013 or LIVE.
As future work, we will study our performance using massively augmented datasets via semi-supervised noisy pseudo-labels.

\noindent\textbf{Acknowledgments}
This work was supported by the Humboldt Foundation.
We thank the organizers of the NTIRE 2022 Perceptual IQA Challenge for their support.


{\small
\bibliographystyle{ieee_fullname}
\bibliography{ref}
}

\end{document}